\documentclass[aps,showpacs,prb,twocolumn,floatfix]{revtex4}
\usepackage{amsmath,amssymb,graphicx,bm}
\begin{document}
\title{Two-particle renormalizations in many-fermion perturbation
  theory: Importance of the Ward identity}

\author{V.  Jani\v{s}}

\affiliation{Institute of Physics, Academy of Sciences of the Czech
  Republic, Na Slovance 2, CZ-18221 Praha 8, Czech Republic}
\email{janis@fzu.cz}

\date{\today}


\begin{abstract}
  We analyze two-particle renormalizations within many-fermion
  perturbation expansion. We show that present diagrammatic theories
  suffer from lack of a direct diagrammatic control over the physical
  two-particle functions. To rectify this we introduce and prove a
  Ward identity enabling an explicit construction of the self-energy
  from a given two-particle irreducible vertex. Approximations
  constructed in this way are causal, obey conservation laws and offer
  an explicit diagrammatic control of singularities in dynamical
  two-particle functions.
\end{abstract}
\pacs{71.10Fd,71.10Li}

\maketitle 
Correlated electrons in metals represent a system with unparalleled
features that remain far from being satisfactorily and fully
understood. They are unique in that respect that to control their
behavior in intermediate and strong coupling regimes we must have at
our disposal effective techniques for a reliable description of
one- and two-particle characteristics. One-electron functions,
such as the spectral function, the self-energy or the density of
states carry information about the fermionic character of elementary
excitations in metals. They are decisive for the Fermi liquid behavior
as well as for deviations from the Landau quasiparticle picture.  The
two-particle correlation and Green functions control the critical
behavior and signal cooperative phenomena and phase transitions via
divergences in generalized susceptibilities. In intermediate coupling,
where the Coulomb repulsion becomes comparable with the kinetic
energy, dynamical fluctuations in the system become strong and may
lead to either a phase transition to a new (magnetic) phase or to a
breakdown of the Fermi-liquid behavior. In this transition regime we
have to treat both single-electron and pair excitations on the same
footing and keep the two-particle functions directly accessible.

One of the most flexible and physical ways to understand various
phenomena of correlated electrons is to use many-body perturbation
theory and Feynman diagrams. Except for the single-impurity Anderson
and Kondo models, where the Fermi-liquid regime survives to infinite
interaction strength,\cite{Yamada75,Zlatic83} we have to renormalize
the perturbation expansion.  Baym and Kadanoff were the first who
showed how to implement renormalizations into the perturbation theory
in a systematic and consistent way.  \cite{Baym61} The basic idea of
their approach is to express physical quantities, and in particular
the self-energy, as a functional of the renormalized one-particle
propagator $G$ and the bare interaction $U$, i.~e., we construct a
self-energy functional $\Sigma_{\sigma}(\mathbf{k},i\omega_n) =
\Sigma_{\sigma}[G,U](\mathbf{k},i\omega_n)$. Once we find an
approximate form of this functional from a diagrammatic expansion free
of self-energy insertions, we add the Dyson equation $
G^{-1}_{\sigma}(\mathbf{k},i\omega_n) =
\mathcal{G}^{-1}_{\sigma}(\mathbf{k},i\omega_n) -
\Sigma_{\sigma}(\mathbf{k},i\omega_n)$
%
to complete our approximation for the self-energy. We used $\mathcal{G}$ to
denote the bare one-particle propagator. If we extend our equations to
situations with external perturbing potentials, we can derive all
necessary thermodynamic functions.\cite{Janis99b} Thermodynamic
consistence and macroscopic conservation laws are thereby guaranteed.

Although we construct a functional for the self-energy from perturbation
theory, the fundamental quantity in the Baym-Kadanoff approach is the
generating functional related to the self-energy via a functional
differential equation
\begin{equation}
  \label{eq:Phi_funct}
  \Sigma_{\sigma}[G,U](\mathbf{k},i\omega_n) = \frac
  {\delta\Phi[G,U]}{\delta G_{\sigma}(\mathbf{k},i\omega_n)}\ .
\end{equation}
When we are able to find the Luttinger-Ward functional $\Phi[G,U]$
explicitly we speak about $\Phi$-derivable
approximations.\cite{Janis99a,Note1}

Higher-order Green functions are derived from the self-energy
functional via functional derivatives that may be viewed upon as
generalized Ward identities. They connect lower-order with higher-order
irreducible (vertex) functions. The identity connecting the
two-particle irreducible vertex with the one-particle one
(self-energy) reads in the direct space
\begin{equation}
  \label{eq:2IP-diff}
  \Lambda_{\sigma\sigma'}^\alpha(13,24)=\frac{\delta\Sigma_{\sigma}(1,2)}
  {\delta_{\alpha} G_{\sigma'}(4,3)} = \frac{\delta^2\Phi[G,U]}
  {\delta_{\alpha} G_{\sigma}(2,1)\delta_{\alpha} G_{\sigma'}(4,3)}
\end{equation}
where we denoted the space-time coordinates $1=(\textbf{R}_1,\tau_1)$ etc.
We also introduced an index $\alpha$ denoting the appropriate
two-particle irreducibility channel. We have three topologically
nonequivalent two-particle irreducibility channels: electron-hole
($\alpha=eh$), electron-electron ($\alpha=ee$), and interaction
($\alpha=U$) channels according to whether we cannot disconnect the
diagram by cutting a pair of an electron and a hole, two electron (hole)
propagators, or by cutting a polarization bubble, respectively, cf.
Ref.~\onlinecite{Janis99a}.  Using functional derivatives of the
generating functional we derive all two-particle and higher-order
Green functions.

The formally exact approach of Baym and Kadanoff uses explicitly only
mass renormalization, i.~e., the self-energy functional depends
explicitly on the renormalized one-electron propagator and the bare
interaction. There is no explicit two-particle renormalization in this
formulation.  The two-particle functions are passive outputs from
functional derivatives of the self-energy functional. We hence cannot
asses or directly control the critical behavior via, for example,
low-energy scalings or summations of most divergent diagrams. Each
change in the two-particle function arises only via an adequate change
in the self-energy, which is too cumbersome and sometimes even not
viable. This is a severe drawback, in particular in critical regions,
where two-particle functions become singular and very sensitive to any
small change.

One can improve upon this by an explicit charge renormalization, that
means that one replaces the bare Coulomb interaction in the
perturbation expansion by appropriate two-particle irreducible
vertices $\Lambda^\alpha$.  We then obtain a new generating
functional in a form $\Phi[G,\Lambda^{\alpha}]$. The most direct way
to do this is to use the so-called parquet approach introduced in the
nonrelativistic many-body theory by De Dominicis and
Martin.\cite{Dominicis62} The parquet approach differs from the
Baym-Kadanoff construction, as standardly used, in that the parquet
scheme takes the two-particle irreducible vertices as primary objects
for which one tries to find a functional representation from the
diagrammatic theory.

Once we have prescriptions for the two-particle irreducible vertices
$\Lambda^{\alpha}$ we use the Bethe-Salpeter equation from the
respective two-particle irreducibility channel to find the full
two-particle vertex
\begin{multline}
  \label{eq:Bethe-Salpeter}
  \Gamma_{\sigma\sigma'}(k;q,q')=\Lambda_{\sigma\sigma'}^\alpha
    (k;q,q')\\ - (1+\delta_{\sigma\sigma'}) \left[\Lambda^\alpha
    GG\odot \Gamma\right]_{\sigma\sigma'} (k;q,q') .
\end{multline}
We introduced four-momenta
$k=(\mathbf{k},i\omega_n),q=(\mathbf{q},i\nu_m)$ for fermionic and
bosonic variables, with Matsubara frequencies $\omega_n=(2n+1)\pi T$
and $\nu_m=2m\pi T$ at temperature $T$. Each two-particle scattering
channel $\alpha$ is characterized by the way the irreducible vertices
are interconnected by pairs of one-particle propagators $GG$, denoted
here by the generic symbol $\odot$.\cite{Janis99b}

Up to this point the one-particle propagator $G$ and the vertices
$\Lambda^{\alpha}$ have been treated as independent. To find a functional
for the self-energy and then also for the generating functional $\Phi$
one uses an exact Schwinger-Dyson equation of motion. Its explicit form
depends on the model we choose. Here we use the lattice Hubbard model
with completely screened, local Coulomb interaction. The
Schwinger-Dyson equation then reads
\begin{multline}
  \label{eq:sigma-2P}
  \Sigma_\sigma(k)=U\sum_{k'} G_{-\sigma}(k')\ -U
  \sum_{k'q}\Gamma_{\sigma-\sigma}(k;q,k'-k)\\ \times
  G_\sigma(k+q)G_{-\sigma}(k'+q)G_{-\sigma}(k').
\end{multline}
We introduced a short-hand notation $\sum_q =
N^{-1}\sum_{\mathbf{q}}\beta^{-1}\sum_{m}$ for $q=(\mathbf{q},i\nu_m)$
and analogously for $k'=(\mathbf{k}',i\omega_{n'})$. The
Schwinger-Dyson equation determines the self-energy functional
$\Sigma_{\sigma}[G,\Lambda^{\alpha}]$ from which we can construct a
generating functional $\Phi[G,\Lambda^{\alpha}]$, which actually has been
done for various two-particle approximations of the parquet
type.\cite{Janis98,Janis99b}
 
In the parquet approach all physical quantities are represented as
functionals of renormalized one-particle propagators $G_{\sigma}$,
representing mass renormalization, and two-particle irreducible
vertices $\Lambda^{\alpha}_{\sigma\sigma'}$, standing for charge
renormalization.  The present way how the parquet-type approximations,
or more generally two-particle renormalizations, are applied, however,
leads to a scheme effectively equivalent to that of Baym and Kadanoff.
An intended direct control over the two-particle functions is lost at
the end.  We use equation \eqref{eq:sigma-2P} to obtain the
self-energy functional in the parquet approach. In conserving
theories the physical two-particle functions must be determined via
functional derivatives as in Eq.~\eqref{eq:2IP-diff}.  The
two-particle vertex constructed via a functional derivative from the
self-energy then differs from the vertex we started with.  Furthermore,
Eq.~\eqref{eq:sigma-2P} does no longer play the role of the
Schwinger-Dyson equation, since the actual physical two-particle
vertex $\Gamma_{phys}$ differs from $\Gamma$ used in
Eq.\eqref{eq:sigma-2P}. This discrepancy reflects the general fact
that we cannot fulfill both Eq.~\eqref{eq:2IP-diff} and
Eq.~\eqref{eq:sigma-2P} unless we have an exact solution for the
two-particle vertex. Hence, Eq.~\eqref{eq:sigma-2P} in the parquet
construction serves as a generator of the self-energy functional from
which all physical quantities are derived via functional differential
equations. The two-particle vertices $\Lambda^\alpha,\Gamma$ we use in
the parquet approach are only auxiliary functions and not the physical
two-particle functions we need.
 
It is clear that in approximate thermodynamically consistent and
conserving schemes in the Baym-Kadanoff form we have to give up the
Schwinger-Dyson equation of motion. A question arises whether also
with two-particle renormalizations we have to lose the initial meaning
and interpretation of the vertex functions. Actually, the two-particle
renormalizations in the parquet approach were introduced in the effort
to gain a better control over the behavior of two-particle functions
in critical regions with singularities in Bethe-Salpeter equations.

To regain the control over the two-particle functions, i.~e., to return the
meaning of two-particle functions to the diagrammatic representations for
the vertices $\Lambda^{\alpha}$, we have to resolve the Ward identity,
Eq.~\eqref{eq:2IP-diff}, for the self-energy. The self-energy then will no
longer be related to the two-particle vertex via Eq.~\eqref{eq:sigma-2P}
but rather through an integral form of the Ward identity. In this way the
two-particle functions will retain their complete diagrammatic
representation enabling a direct control of their critical behavior. Using
a Ward identity for the determination of the self-energy from a given
two-particle vertex guarantees thermodynamic consistence of the theory.

\begin{figure}
  \centering
  \includegraphics[height=27mm]{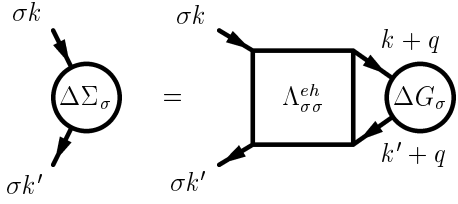}
  \caption{\label{fig:Ward}Ward identity connecting the self-energy
    and the triplet irreducible vertex. The interaction acts
    vertically between the upper and the lower line in two-particle
    functions.}
\end{figure}

The aim of this paper is to prove the following partially integrated
Ward identity between the self-energy and the electron-hole
irreducible triplet vertex
\begin{multline}
    \label{eq:WI-integral}
    \Sigma_{\sigma}(k) - \Sigma_{\sigma}(k') = \sum_{q}
    \Lambda^{eh}_{\sigma\sigma} (k;q,k'-k)\\ \times
    \left[G_{\sigma}(k+q) - G_{\sigma}(k'+q)\right]\ .
\end{multline}
Its diagrammatic representation is given in Fig.~\ref{fig:Ward}.  Equation
\eqref{eq:WI-integral} holds for arbitrary four-momenta $k$ and $k'$. It is
a generalization of a Ward identity from noninteracting electrons
in a random potential proved by Vollhardt and W\"olfle. \cite{Vollhardt80}

We stress that the Ward identity \eqref{eq:WI-integral} can be proved
only for the triplet irreducible vertex $\Lambda_{\sigma\sigma}$.
This is important in particular for the Hubbard model where the local
Coulomb interaction acts only between particles with opposite spins
and singlet vertex functions $\Lambda^{\alpha}_{\sigma-\sigma}$ are
preferably used.

\begin{figure}
\includegraphics[height=27mm]{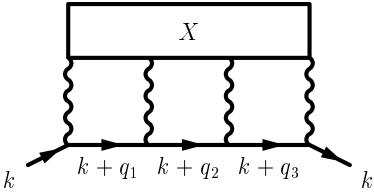}
  \caption{A typical self-energy diagram with internal loops ($X$)
  connected with the propagating electron only via the
  interaction (wavy) lines.}
  \label{fig:Sigma}
\end{figure}

We use similar diagrammatic arguments as applied in noninteracting
disordered systems to prove identity \eqref{eq:WI-integral}. Due to
spin and charge conservation in the vertices of the perturbation
theory, each diagram for the self-energy contains just a single
electron trajectory connecting the incoming with the outgoing external
electron line. We set aside this fundamental fermion trajectory
propagating the incoming charge and spin from the rest of the diagram
that we denote $X$. Function $X$ contains interaction lines and only
closed loops of fermion propagators. It is connected with the
fundamental fermion trajectory via interactions as shown in
Fig.~\ref{fig:Sigma}.  Each self-energy diagram can be classified
according to the length (number of scattering events) of the
fundamental fermion trajectory.

We do not need to consider the Hartree term, since it does not
contribute to the r.h.s. of Eq.~\eqref{eq:WI-integral}. Hence the
dynamical self-energy can be represented by an expansion
\begin{multline}
    \label{eq:Sigma-decoupling}
    \Sigma_{\sigma}(k) = \sum_{n=1}^{\infty} U^{n+1}\\ \times
    \sum_{q_1,\ldots,q_n}X^{(n)}[G,U]
    (q_1,q_2-q_1,\ldots,q_{n}-q_{n-1}, -q_n)\\
    \times G_{\sigma}(k+q_1) G_{\sigma}(k+q_2) \ldots
    G_{\sigma}(k+q_n)
\end{multline}
where $n$ denotes the length of the fundamental fermion trajectory
($n=3$ in Fig.~\ref{fig:Sigma}). Note that the expansion on the r.h.s.
of Eq.~\eqref{eq:Sigma-decoupling} is not an expansion in the
interaction strength, since the loop function $X$ contains
interaction. It is clear that the loop function depends only on the
transfer (bosonic) momenta $q_1,\ldots,q_n$ and not on the incoming
fermionic momentum $k$.

Only the one-electron propagators contributing to the fundamental
fermion trajectory are relevant for the proof of Eq.~\eqref{eq:WI-integral},
since only there we can distinguish different external momenta on the
l.h.s. of Eq.~\eqref{eq:WI-integral}. We utilize the following
identity
\begin{multline}
    \label{eq:G-difference}
    G_{\sigma}(k+q_1)\ldots G_{\sigma}(k+q_n) -
    G_{\sigma}(k'+q_1)\ldots  G_{\sigma}(k'+q_n)\\
    =\sum_{i=1}^n G_{\sigma}(k+q_1)\ldots G_{\sigma}(k+q_{i-1})
    \left[G_{\sigma}(k+q_i) \right. \\
    \left. -\ G_{\sigma}(k'+q_i)\right] G_{\sigma}(k'+q_{i+1})\ldots
    G_{\sigma}(k'+q_{n})
\end{multline}
and rewrite the expansion for the self-energy to
\begin{widetext}
  \begin{multline}
    \label{eq:Sigma-Lambda}
   \Sigma_{\sigma}(k) - \Sigma_{\sigma}(k') =  \sum_{n=1}^{\infty}
    U^{n+1}\sum_{i=1}^{n} \sum_{q_1,\ldots,q_n}
    X^{(n)}(q_1,q_2-q_1,\ldots,q_{n}-q_{n-1},-q_n)\\
    \times G_{\sigma}(k+q_1)\ldots G_{\sigma}(k+q_{i-1})
    G_{\sigma}(k'+q_{i+1})\ldots G_{\sigma}(k'+q_{n})
    \left[G_{\sigma}(k+q_i) - G_{\sigma}(k'+q_i)\right] \\
     =  \sum_{n=1}^{\infty}
     U^{n+1}\sum_{i=0}^{n-1}\Lambda_{\sigma\sigma}^{(i,n-i-1)}(k;q,k'-k)
     \left[G_{\sigma}(k+q) - G_{\sigma}(k'+q)\right] \ .
  \end{multline}
\end{widetext}
We denoted contributions to an electron-hole vertex with $i$ electron
and $n-i$ hole propagators from the fundamental fermion lines of the
two-particle function as $\Lambda_{\sigma\sigma}^{(i,n-i)}$.  The
vertex $\Lambda$ has the electron-hole structure, since the
propagation in momentum $k'$ is in the opposite direction to the
propagation of momentum $k$. The vertex $\Lambda$ must be irreducible
in the electron-hole channel, since all reducible terms are already
incorporated in the renormalized one-electron propagators. On the
other hand, each $eh$-irreducible two-particle diagram can be uniquely
closed to a self-energy via the fundamental fermion trajectory. The
expansion in $n$ and $i$ hence covers all contributions to the
irreducible electron-hole vertex and we can write\cite{Note2}
\begin{equation}
  \label{eq:Lambda-decoupling}
  \Lambda_{\sigma\sigma}^{eh}(k;q,q') = U \sum_{n=0}^{\infty}
  U^{n+1}\sum_{i=0}^{n}\Lambda_{\sigma\sigma}^{(i,n-i)}(k;q,q')\ .
\end{equation}
Inserting Eq.~~\eqref{eq:Lambda-decoupling} into
Eq.~\eqref{eq:Sigma-Lambda} we reveal identity \eqref{eq:WI-integral}.

Relation \eqref{eq:WI-integral} has two important consequences. First,
continuity equation for two-particle correlation functions can be
proved. Second, Eq.~\eqref{eq:WI-integral} can be used to replace
Eq.~\eqref{eq:sigma-2P} in the determination of the self-energy from a
given two-particle vertex $\Lambda^{eh}_{\sigma\sigma}$.
 
To prove a two-particle continuity equation we use the following relation
\begin{multline}
  \label{eq:GG-product}
  G_{\sigma}(k)G_{\sigma}(k+q)\\ = \frac {G_{\sigma}(k) -
    G_{\sigma}(k+q)} {i\nu_m -\epsilon(\mathbf{k}+\mathbf{q}) +
    \epsilon(\mathbf{k}) - \Sigma_{\sigma}(k+q) + \Sigma_{\sigma}(k)}
\end{multline}
in the Bethe-Salpeter equation for the two-particle function
$L_{\sigma\sigma}(k;q,q') = G_{\sigma}(k) G_{\sigma}(k+q')[\delta(q) +
\Gamma_{\sigma\sigma}(k;q,q')G_{\sigma}(k+q) G_{\sigma}(k+q+q')$. The
four-momentum delta function for $q=(\mathbf{q},i\nu_m)$ reads
$\delta(q)=N\delta_{\mathbf{q},\mathbf{0}} \beta\delta_{m,0}$.
Function $L_{\sigma\sigma}$ is the two-particle Green function from
which the exchange term was removed so that it fulfills a
Bethe-Salpeter equation.  If we further utilize the symmetry of the
two-particle vertex we find another form of the Ward identity $\sum_k
\Delta_q G_{\sigma} \Lambda_{\sigma\sigma}^{eh}(k;k'-k,q) =
\Delta_q\Sigma_{\sigma}(k')$.  When we multiply the Bethe-Salpeter
equation for $L_{\sigma\sigma}$ in the electron-hole channel by the
denominator of Eq.~\eqref{eq:GG-product} and integrate over the
incoming and outgoing fermionic momenta we obtain a continuity
equation having in the limit of small transfer three-momenta
$\mathbf{q}$ the following form
\begin{equation}
  \label{eq:continuity}
  i\nu_m \Xi_{\sigma\sigma}(\mathbf{q},i\nu_m) -
  \mathbf{q}\cdot\mathbf{\Xi}^{v}_{\sigma\sigma}(\mathbf{q},i\nu_m)  = 0 .
\end{equation}
We denoted $\Xi_{\sigma\sigma}(\mathbf{q},i\nu_m) = \sum_{k,q'}
L_{\sigma\sigma}(k;q',q)$ and
$\mathbf{\Xi}^v_{\sigma\sigma}(\mathbf{q},i\nu_m) = \sum_{k,q'}
\mathbf{\nabla}\epsilon(\mathbf{k}) L_{\sigma\sigma}(k;q',q)$.
 
To use Eq.~\eqref{eq:WI-integral} in the determination of the
self-energy, we put $\mathbf{k}'=\mathbf{k}$ but let the Matsubara
frequencies $i\omega_n$ and $i\omega_{n'}$ independent so that we can
analytically continue the self-energy difference to the case with
$\omega+i\eta$ and $\omega-i\eta$. We then use an analytically
continued vertex function
$\Lambda_{\sigma\sigma}^{eh}(\mathbf{k},\omega+i\eta;
\mathbf{q},\zeta,\mathbf{0},-2i\eta)$ in the Ward identity
\eqref{eq:WI-integral} to construct the imaginary part of the
self-energy $\Sigma_\sigma(\mathbf{k},\omega+i\eta)$ along the real
axis. The intermediate complex frequency $\zeta$ takes values from an
integration contour used to replace the sum over Matsubara
frequencies. The shape of the integration contour depends on the
analytic structure of the approximate vertex.  The real part of the
self-energy is calculated from the Kramers-Kronig relation
acomplishing thus a causal theory.  This construction of the
self-energy from the vertex function was already successfully applied
in the parquet approach to noninteracting disordered electron
systems.\cite{Janis01b}

To conclude, the principal result of the paper is a partially
integrated Ward identity, Eq.~\eqref{eq:WI-integral}. It is a
consequence of charge and spin conserving particle interaction and was
proved by means of a diagrammatic expansion.  The primary importance
of identity \eqref{eq:WI-integral} lies in the possibility to regain
a direct diagrammatic control over the (critical) behavior of
dynamical two-particle functions by defining the self-energy from the
triplet vertex function via Eq.~\eqref{eq:WI-integral}.  The present
schemes with two-particle (vertex) renormalizations do not obey the
Ward identity \eqref{eq:WI-integral}, except for its infinitesimal
limit, and hence are unable to determine the self-energy from the
physical two-particle vertices. Utilization of the Ward identity
\eqref{eq:WI-integral} opens new possibilities to study critical
behavior of correlated electrons, in particular, when we subject the
electrons to a random potential and investigate the effects of
electron interactions on the metal-insulator
transition.

The work was supported in part by Grant No. 202/01/0764 of the Grant
Agency of the Czech Republic and during the author's stay at the
workshop ``Realistic Theories of Correlated Electron Materials'' at
KITP UCSB, where the research was completed, 
by US NSF Grant No. PHY99-07949.

\end{document}